\documentclass{ws-book9x6}

\usepackage{ws-book-har}           

\title{My Book Title}              

\makeindex

\begin{document}

\titlepages                        





\tableofcontents



\setcounter{page}{1}




%
%
\chapter{Adaptive Space-Time Beamforming in Radar
Systems} \label{chrdl}

\section{Introduction}\label{sec1.1}

Space-time adaptive processing (STAP) techniques
\cite{klemm},\cite{melvin} have been thoroughly investigated in the
last decades as a key enabling technology for advanced airborne
radar applications following the seminal work by Brennan and Reed
\cite{brennan}. A great deal of attention has been given to STAP
algorithms and different strategies to design space-time beamformers
to mitigate the effect of clutter and jamming signals \cite{reed}-
\cite{guerci}. It is well understood that STAP techniques can
improve slow-moving target detection through better mainlobe clutter
suppression, provide better detection in combined clutter and
jamming environments, and offer a significant increase in output
signal to- interference-plus-noise-ratio (SINR). Moreover, it is
also well understood that clutter and jamming signals often reside
in a low-rank signal subspace, which is typically much lower than
the number of degrees of freedom of the array and the associated
space-time beamformer. Due to the large computational complexity of
the matrix inversion operation, the optimum STAP processor is
prohibitive for practical implementation. In addition, another very
challenging issue that is encountered by the optimal STAP technique
is when the number of elements $M$ in the spatio-temporal beamformer
is large. It is well known that $K \geq 2M$ independent and
identically distributed (i.i.d) training samples are required for
the beamformer to achieve the steady-state performance
\cite{haykin}. Thus, in dynamic scenarios the optimal STAP with
large $M$ usually fails or provides poor performance in tracking
target signals contaminated by interference and noise.

In the recent years, a number of innovative space-time beamforming
algorithms have been reported in the literature for clutter and
interference mitigation in radar systems. These algorithms include
low-rank and reduced-dimension techniques
\cite{haimovich}-\cite{jidf}, which employ a two-stage processing
framework to exploit the low-rank property of the clutter and the
jamming signals. The first stage performs dimensionality reduction
and is followed by a second stage that employs a beamforming
algorithm with a reduced dimensional filter. Another class of
important space-time beamforming algorithms adopt the strategy of
compressive sensing and sparsity-aware algorithms, which exploit the
fact that space-time beamformers do not need all their degrees of
freedom to mitigate clutter and jamming signals. These algorithms
compute sparse space-time beamformers which can converge faster and
are effective for STAP in radar systems. By exploiting the low-rank
properties of the interference and devising sparse STAP algorithms,
designers make use of prior knowledge about the clutter and the
jamming signals. It has been recently shown that it is beneficial in
terms of performance to also exploit prior knowledge about the
environment and the data in the form of a known covariance data
matrix. The class of space-time beamforming algorithms that exploit
different forms of prior knowledge are called knowledge-aided STAP
(KA-STAP) algorithms.

The goal of this chapter is to review the recent work and advances
in the area of space-time beamforming algorithms and their
application to radar systems. These systems include phased-array
\cite{melvin} and multi-input multi-output (MIMO) radar systems
\cite{haimo_08}, mono-static and bi-static radar systems and other
configurations \cite{melvin}. Furthermore, this chapter also
describes in detail some of the most successful space-time
beamforming algorithms that exploit low-rank and sparsity properties
as well as the use of prior-knowledge to improve the performance of
STAP algorithms in radar systems.

The chapter is structured as follows. Section \ref{sec1} describes
the radar system under consideration and the signal model used to
mathematically describe it. Section \ref{sec2} formulates the
problem of designing space-time beamformers and reviews conventional
space-time beamforming algorithms. Section \ref{sec3} examines
low-rank space-time beamforming algorithms, whereas Section
\ref{sec4} explores the concept of sparsity-aware space-time
beamforming algorithms. Section \ref{sec5} studies knowledge-aided
beamforming algorithms and discusses how these techniques can be
adopted in existing radar systems. Section \ref{sec6} is devoted to
the presentation of simulation results, discussions and the
comparison of a number of existing algorithms. The chapter ends with
Section \ref{sec7} which gives the concluding remarks of this
chapter.

%
%
%
%
%
%
%

\section{System and Signal Models}\label{sec1}

\begin{figure}[!htb]
\begin{center}
\def\epsfsize#1#2{1\columnwidth}
\epsfbox{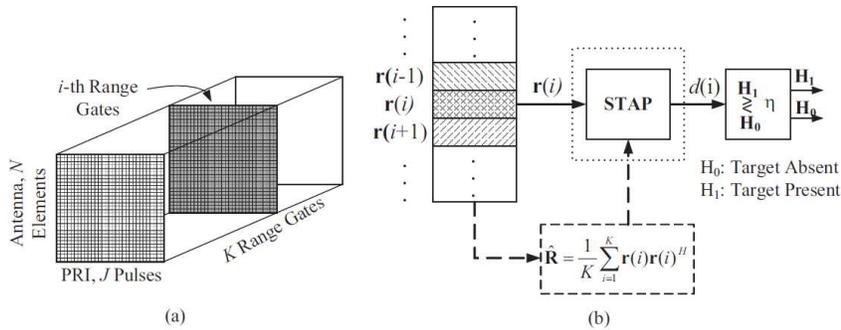} \caption{(a) The Radar CPI datacube. (b) The STAP
schematic.} \label{system}
\end{center}
\end{figure}

The system under consideration is a pulsed Doppler radar residing on
an airborne platform. The radar antenna is a uniformly spaced linear
antenna array consisting of $N$ elements. The radar returns are
collected in a coherent processing interval (CPI), which is referred
to as the $3$-D radar datacube shown in Fig. \ref{system} (a), where
$K$ denotes the number of samples collected to cover the range
interval. The data is then processed at one range of interest, which
corresponds to a slice of the CPI datacube. This slice is a $J
\times N$ matrix which consists of $N \times 1$ spatial snapshots
for $J$ pulses at the range of interest. It is convenient to stack
the matrix column-wise to form the $M \times 1$ vector r(i), termed
the i-th range gate spacetime snapshot, where $M=JN$ and  $1 < i
\leq K$ \cite{klemm}.

The objective of a radar is to ascertain whether targets are present
in the data. Thus, given a space-time snapshot, radar detection is a
binary hypothesis problem, where hypothesis $H_0$ corresponds to the
absence of a target and hypothesis $H_1$ corresponds to the presence
of a target. The radar space-time snapshot is then expressed for
each of the two hypotheses in the following form
\begin{equation}
\begin{split}
H_0 : {\boldsymbol r}(i) & = {\boldsymbol v}(i);\\
H1 : {\boldsymbol r}(i) & = a{\boldsymbol s} + {\boldsymbol v}(i);
\end{split}
\end{equation}
where $a$ is a zero-mean complex Gaussian random variable with
variance $\sigma_s^2$ , ${\boldsymbol v}(i) = {\boldsymbol r}_c(i)+
{\boldsymbol r}_j(i) + {\boldsymbol n}(i)$ contains the input
interference-plus-noise vector which consists of the clutter
${\boldsymbol r}_c(i)$, the jamming signal ${\boldsymbol r}_j(i)$
and the complex white Gaussian noise ${\boldsymbol n}(i)$. These
three components are assumed to be mutually uncorrelated. Thus, the
$M \times M$ covariance matrix ${\boldsymbol R}$ of the undesired
clutter-plus-jammer-plus-noise component can be modelled as
\begin{equation}
{\boldsymbol R} = {\boldsymbol R}_c + {\boldsymbol R}_j +
{\boldsymbol R}_n
\end{equation}
where $(\cdot)^H$ represents the Hermitian transpose and $E[ \cdot]$
denotes expectation. The noise covariance matrix is given by
${\boldsymbol R}_n = E[{\boldsymbol n}(i) {\boldsymbol n}^H(i)] =
\sigma_n^2 {\boldsymbol I} $, where $\sigma_n^2$ is the variance of
the noise and ${\boldsymbol I}$ is an identity matrix.  The clutter
signal can be modeled as the superposition of a large number of
independent clutter patches which are evenly distributed in azimuth
about the receiver. Thus, the clutter covariance matrix can be
expressed as
\begin{equation}
{\boldsymbol R}_c = E[{\boldsymbol r}_c{\boldsymbol r}_c^H] =
\sum_{k=1}^{N_r}\sum_{l=1}^{N_c} \xi_{k,l}^c [{\boldsymbol
b}(\vartheta_{k,l}^c){\boldsymbol b}(\vartheta_{k,l}^{c,~H}) \otimes
[{\boldsymbol a}(\varpi_{k,l}^c){\boldsymbol
a}(\varpi_{k,l}^{c,~H}),
\end{equation}
where $N_r$ denotes the number of range ambiguities and $N_c$
denotes the number of clutter patches. The quantity $\xi^c_{k,l}$ is
the power of the reflected signal by the $k,l$-th clutter patch. The
symbol $\otimes$ denotes Kronecker product, and the quantities
${\boldsymbol b}(\vartheta_{k,l}^{c}$ and ${\boldsymbol
a}(\varpi_{k,l}^c)$ denote the spatial steering vector with the
spatial frequency $\vartheta^c_{k,l}$ and the temporal steering
vector with the normalized Doppler frequency $\varpi^c_{k,l}$ for
the $k,l$-th clutter patch, respectively, which can be expressed as
follows
\begin{equation}
{\boldsymbol b}(\vartheta_{k,l}^c) = \left[\begin{array}{c} \\ 1 \\
e^{-j2\pi \vartheta} \\  e^{-j2\pi 2 \vartheta}\\ \vdots \\
e^{-j2\pi (N-1) \vartheta} \end{array} \right],~~ {\boldsymbol a}(\varpi_{k,l}^c) = \left[\begin{array}{c} \\ 1 \\
e^{-j2\pi \varpi} \\  e^{-j2\pi 2 \varpi}\\ \vdots \\
e^{-j2\pi (N-1) \varpi} \end{array} \right],
\end{equation}
where $\vartheta = \frac{d}{\lambda} cos(\phi) sin(\theta)$ and $
\vartheta= f_d/f_r$, $\lambda$ is the wavelength, $d$ is the
inter-element spacing which is normally set to half wavelength, and
$\phi$ and $\theta$ are the elevation and the azimuth angles,
respectively. The quantities $f_d$ and $f_r$ are the Doppler
frequency and the pulse repetition frequency, respectively. The
jamming covariance matrix ${\boldsymbol R}_j = E[{\boldsymbol
r}_j(i){\boldsymbol r}^H_j(i)]$ can be written as
\begin{equation}
{\boldsymbol R}_j =  \sum_{q=1}^{N_j} \xi_q^j [{\boldsymbol
b}(\vartheta_q^j) {\boldsymbol b}^H(\vartheta_q^j)] \otimes
{\boldsymbol I}_K,
\end{equation}
where $\xi_q^j$ is the power of the $q$-th jammer. The vector
${\boldsymbol b}(\vartheta_q^j)$ is the spatial steering vector with
the spatial frequency $\vartheta_q^j$ of the $q$-th jammer and $N_j$
is the number of jamming signals. The vector ${\boldsymbol s}$ is
the $M \times 1$ normalized space-time steering vector in the
space-time look-direction, which can be defined as
\begin{equation}
{\boldsymbol s} = \sqrt{\xi_t} {\boldsymbol b}(\vartheta_t)
{\boldsymbol a} (\varpi_t),
\end{equation}
where ${\boldsymbol a} (\varpi_t)$ is the $K \times 1$ normalized
temporal steering vector at the target Doppler frequency $\varpi_t$
and $b(\vartheta_t)$ is the $N \times 1$ normalized spatial steering
vector in the direction provided by the target spatial frequency
$\vartheta_t$ and $\xi_t$ denotes the power of the target.

%

\section{Conventional Beamforming Algorithms} \label{sec2}

In order to detect the presence of targets, each range bin is
processed by an adaptive space-time beamformer, which is typically
designed to achieve maximum output SINR, followed by a hypothesis
test to determine the target presence or absence. The secondary data
${\boldsymbol r}(i)$ are taken from training samples, which should
be ideally i.i.d. training samples but are often non-heterogeneous
\cite{klemm}. The optimum full-rank STAP that maximizes the SINR can
obtained by solving the following minimum variance distortionless
response (MVDR) constrained optimization given by:
\begin{equation}
{\boldsymbol w}_{\rm opt} = \arg \min_{\boldsymbol w} {\boldsymbol
w}^H {\boldsymbol R} {\boldsymbol w} ~~~{\rm subject}~{\rm to}~
{\boldsymbol w}^H {\boldsymbol s} = 1,
\end{equation}
where the optimal space-time MVDR beamformer ${\boldsymbol w}_{\rm
opt}$ is designed to maximize the SINR and to maintain a normalized
response in the target spatial-Doppler look-direction. The solution
to the optimization problem above is described by:
\begin{equation}
{\boldsymbol w}_{\rm opt} = \frac{{\boldsymbol R}^{-1}{\boldsymbol
s}}{{\boldsymbol s}^H {\boldsymbol R}^{-1}{\boldsymbol s}}.
\label{conv_bf}
\end{equation}
The space-time beamformer ${\boldsymbol w}_{\rm opt}$ can be
computed by using the above solution. Alternatively, the space-time
beamformer can be estimated using adaptive algorithms \cite{haykin}.
These algorithms include the least mean-square (LMS), the conjugate
gradient (CG) and the recursive least-squares (RLS) techniques. The
computational complexity of these algorithms ranges from linear with
$M$ for the LMS to quadratic with $M$ for the CG and RLS algorithms.
A common problem with the conventional adaptive algorithms is that
the laws that govern their convergence and tracking behaviors imply
that they depend on $M$ and on the eigenvalue spread of
${\boldsymbol{R}}$. This means that their performance may degrade
significantly when the space-time beamformer has many parameters for
adaptation, which makes the computation of the parameters of the
beamformer slow and costly. This problem can be addressed by some
recent techniques reported in the literature, namely, low-rank,
sparsity-aware and knowledged-aided algorithms.

\section{Low-Rank Beamforming Algorithms} \label{sec3}

Low-rank adaptive signal processing has been considered as a key
technique for dealing with large systems in the last decade. The
basic idea of the low-rank algorithms is to reduce the number of
adaptive coefficients by projecting the received vectors onto a
lower dimensional subspace which consists of a set of basis vectors.
The adaptation of the low-order filter within the lower dimensional
subspace results in significant computational savings, faster
convergence speed and better tracking performance. The first
statistical low-rank method was based on a principal-components (PC)
decomposition of the target-free covariance matrix \cite{haimovich}.
Another class of eigen-decomposition methods was based on the
cross-spectral metric (CSM) \cite{goldstein97,goldstein}. Both the
PC and the CSM algorithms require a high computational cost due to
the eigen-decomposition. A family of the Krylov subspace methods has
been investigated thoroughly in the recent years. This class of
low-rank algorithms includes the multistage Wiener filter (MSWF)
\cite{goldstein98,guerci,gau} which projects the observation data
onto a lower-dimensional Krylov subspace, and the auxiliary-vector
filters (AVF) \cite{pados99,pados01,pados07}. These methods are
relatively complex to implement in practice and may suffer from
numerical problems despite their improved convergence and tracking
performance. The joint domain localized (JDL) approach, which is a
beamspace reduced-dimension algorithm, was proposed by Wang and Cai
\cite{jdl} and investigated in both homogeneous and nonhomogeneous
environments in \cite{adve1,adve2}, respectively. Recently,
reduced-rank adaptive processing algorithms based on joint iterative
optimization of adaptive filters
\cite{delamare_spl07,fa,delamareel,delamare_esp,fa2011} and based on
an adaptive diversity-combined decimation and interpolation scheme
\cite{delamare_icassp07,jidf,RuiSTAP2010,barc} were proposed,
respectively.

\begin{figure}[!htb]
\begin{center}
\def\epsfsize#1#2{1\columnwidth}
\epsfbox{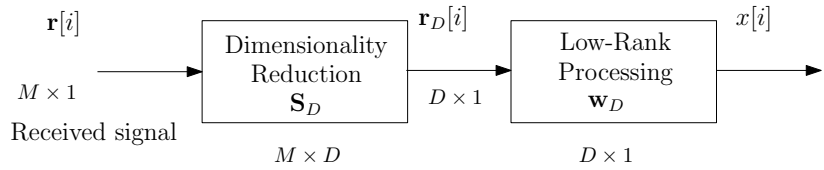} \vspace*{-1em}\caption{\small Low-rank signal
processing scheme.} \label{fig2}
\end{center}
\end{figure}

The basic idea of low-rank algorithms is to reduce the number of
adaptive coefficients by projecting the received vectors onto a
lower dimensional subspace. Let ${\boldsymbol S}_D$ denote the $M
\times D$ rank-reduction matrix with column vectors which form an $M
\times 1$ basis for a $D$-dimensional subspace, where $D < M$. Thus,
the received signal ${\boldsymbol r}(i)$ is transformed into its
reduced-rank version ${\boldsymbol r}_D(i)$ given by
\begin{equation}
{\boldsymbol r}_D(i) = {\boldsymbol S}_D^H {\boldsymbol r}(i)
\end{equation}
The low-rank signal is processed by an adaptive low-rank space-time
beamformer ${\boldsymbol w}_D$ with $D$ coefficients. This is
illustrated in Fig. 2. Subsequently, the decision is made based on
the output of the beamformer $y(i) = {\boldsymbol w}_D^H
{\boldsymbol S}_D^H{\boldsymbol r}(i)$. A designer can compute the
parameters of the beamformer by solving the following constrained
optimization problem:
\begin{equation}
{\boldsymbol w}_{D,{\rm opt}} = \arg \min_{{\boldsymbol w}_D}
{\boldsymbol w}^H_D {\boldsymbol S}_D^H{\boldsymbol R}{\boldsymbol
S}_D {\boldsymbol w}_D ~~~{\rm subject}~{\rm to}~ {\boldsymbol
w}^H_D{\boldsymbol S}_D^H {\boldsymbol s} = 1,
\end{equation}
The optimal low-rank MVDR solution for the above problem is given by
\begin{equation}
{\boldsymbol w}_{D,{\rm opt}} = \frac{({\boldsymbol
S}_D^H{\boldsymbol R}{\boldsymbol S}_D)^{-1} {\boldsymbol S}_D^H
{\boldsymbol s}}{{\boldsymbol s}^H{\boldsymbol S}_D ({\boldsymbol
S}_D^H{\boldsymbol R}{\boldsymbol S}_D )^{-1}{\boldsymbol S}_D^H
{\boldsymbol s}} =  \frac{{\boldsymbol R}_D^{-1} {\boldsymbol
s}_D}{{\boldsymbol s}^H_D{\boldsymbol R}_D^{-1}{\boldsymbol s}_D}.
\end{equation}
where ${\boldsymbol R}_D =  {\boldsymbol S}_D^H{\boldsymbol
R}{\boldsymbol S}_D$ denotes the low-rank covariance matrix and
${\boldsymbol s}_D = {\boldsymbol S}_D^H {\boldsymbol s}$ denotes
the low-rank steering vector. The key challenge in the design of
low-rank STAP algorithms is to find a cost-effective method to
compute the rank-reduction matrix ${\boldsymbol S}_D$.

\subsection{ Eigenvalue-decomposition-based algorithms}

The eigenvalue-decomposition (EVD)-based beamforming algorithms are
also known as PC-based algorithms and have been originally reported
as the eigencanceler method. These PC-based algorithms refer to the
beamformers constructed with a subset of the eigenvectors of the
interference-only covariance matrix associated with the eigenvalues
of largest magnitude. The first application of this method to radar
systems was reported in \cite{haimovich}.

The basic idea of the EVD-based beamformer is to approximate the $M
\times M$ covariance matrix ${\boldsymbol R}$ of the received data
as follows:
\begin{equation}
{\boldsymbol R} = \sum_{d=1}^{D} \lambda_d {\boldsymbol v}_d
{\boldsymbol v}_d^H,
\end{equation}
where the $M \times 1$ vector ${\boldsymbol v}_d$ corresponds to the
$d$th eigenvector of ${\boldsymbol R}$ and $\lambda_d$ is the $d$th
eigenvalue of ${\boldsymbol R}$. By assuming that the eigenvalues
are obtained in decreasing order of magnitude, the EVD-based method
approximates ${\boldsymbol R}$ using its $D$ dominant eigenvectors.
The rank-reduction matrix is constructed by using the $D$ dominant
eigenvectors as described by
\begin{equation}
{\boldsymbol S}_D = [{\boldsymbol v}_1~ {\boldsymbol v}_2 ~\ldots
~{\boldsymbol v}_D]
\end{equation}
The low-rank MVDR solution for the above problem is given by
\begin{equation} {\boldsymbol w}_{D} = \frac{{\boldsymbol S}_D({\boldsymbol
S}_D^H{\boldsymbol R}{\boldsymbol S}_D)^{-1} {\boldsymbol S}_D^H
{\boldsymbol s}}{{\boldsymbol s}^H{\boldsymbol S}_D ({\boldsymbol
S}_D^H{\boldsymbol R}{\boldsymbol S}_D )^{-1}{\boldsymbol S}_D^H
{\boldsymbol s}} = \frac{(\sum_{d=1}^{D} \lambda_d^{-1} {\boldsymbol
v}_d {\boldsymbol v}_d^H) {\boldsymbol s}}{{\boldsymbol s}^H
(\sum_{d=1}^{D} \lambda_d^{-1} {\boldsymbol v}_d {\boldsymbol
v}_d^H){\boldsymbol s}}.
\end{equation}
The EVD-based low-rank MVDR space-time beamformer described above
does not take into account the target steering vector ${\boldsymbol
s}$ when selecting a suitable subspace representation of the
interference. Clearly, this low-rank space-time beamformer requires
the computation of an EVD, which has a computational cost that is
cubic with $M$ \cite{golub}. In order to reduce this computational
complexity, a designer can resort to subspace tracking algorithms
which bring the cost down to $O(M^2)$ \cite{Yang,Badeau}. Another
technique associated with EVD-based beamforming that can improve the
performance of low-rank MVDR space-time beamformers is the method
called cross-spectral metric (CSM) \cite{goldstein97}. The CSM
approach chooses the set of $D$ eigenvectors for the rank-reduction
matrix which optimizes the desired criterion, namely, the
maximization of the SINR, in opposition to the PC method which
always chooses the dominant eigenvectors.

\subsection{Krylov subspace-based algorithms}

The first Krylov methods, namely, the conjugate gradient (CG) method
\cite{cga} and the Lanczos algorithm \cite{lanczos} have been
originally proposed for solving large systems of linear equations.
These algorithms used in numerical linear algebra are mathematically
identical to each other and have been derived for Hermitian and
positive definite system matrices. Other techniques have been
reported for solving these problems and the Arnoldi algorithm
\cite{arnoldi} is a computationally efficient procedure for
arbitrarily invertible system matrices. The multistage Wiener filter
(MSWF) \cite{goldstein97} and the auxiliary vector filtering (AVF)
\cite{pados99} algorithms are based on a multistage decomposition of
the linear MMSE estimator. A key feature of these methods is that
they do not require an EVD and have a very good performance. It
turns out that Krylov subspace algorithms that are used for solving
very large and sparse systems of linear equations, are highly
suitable alternatives for designing low-rank space-time beamforming
algorithms in radar systems. The basic idea of Krylov subspace
algorithms is to construct the rank-reduction matrix ${\mathbf S}_D$
with the following structure:
\begin{equation}
{\boldsymbol S}_{D} = \big[ {\boldsymbol q} ~ {\boldsymbol R}
{\boldsymbol q} ~\ldots ~ {\boldsymbol R}^{D-1} {\boldsymbol q}
\big], \label{kry}
\end{equation}
where ${\boldsymbol q} = \frac{{\boldsymbol s}}{||{\boldsymbol
s}[i]||}$ and $||\cdot||$ denotes the Euclidean norm (or the
$2$-norm) of a vector. In order to compute the basis vectors of the
Krylov subspace (the vectors of ${\boldsymbol S}_{D}$), a designer
can either directly employ the expression in (\ref{kry}) or resort
to more sophisticated approaches such as the Arnoldi iteration
\cite{arnoldi}. The low-rank MVDR solution for the space-time
beamformer using the Krylov subspace is given by
\begin{equation}
{\boldsymbol w}_{D} = \frac{({\boldsymbol
S}_D^H{\boldsymbol R}{\boldsymbol S}_D)^{-1} {\boldsymbol S}_D^H
{\boldsymbol s}}{{\boldsymbol s}^H{\boldsymbol S}_D ({\boldsymbol
S}_D^H{\boldsymbol R}{\boldsymbol S}_D )^{-1}{\boldsymbol S}_D^H
{\boldsymbol s}} .
\end{equation}
An appealing feature of the Krylov subspace algorithms is that the
required model order $D$ does not scale with the system size.
Indeed, when $M$ goes to infinity the required $D$ remains a finite
and relatively small value. This result was established in
\cite{xiao}. Among the disadvantages of Krylov subspace methods are
the relatively high computational cost of constructing ${\boldsymbol
S}_{D}$ ($O(DM^2)$), the numerical instability of some
implementations and the lack of flexibility for imposing constraints
on the design of the basis vectors.

\subsection{JIO-based algorithms}

The aim of this part is to introduce the reader to low-rank
beamforming algorithms based on joint iterative optimization (JIO)
techniques. The idea of these methods is to design the main
components of a low-rank space-time beamforming scheme via a general
optimization approach. The basic ideas of JIO techniques have been
reported in \cite{delamare_spl07,fa,delamareel,delamare_esp,fa2011}.
Amongst the advantages of JIO techniques are the flexibility to
choose the optimisation algorithm and to impose constraints, which
provides a significant advantage over eigen-based and Krylov
subspace methods. One disadvantage that is shared amongst the JIO
techniques, eigen-based and Krylov subspace methods are the
complexity associated with the design of the matrix ${\mathbf
S}_{D}$. For instance, if we are to design a beamforming algorithm
with a very large $M$, we still have the problem of having to design
an $M \times D$ rank-reduction matrix ${\mathbf S}_{D}$.

In the framework of JIO techniques, the design of the matrix
${\mathbf S}_{D}$ and the beamforming vector ${\mathbf w}_D$ for a
fixed model order $D$ will be dictated by the optimization problem
and the algorithm chosen to compute the solution. To this end, we
will focus on a generic ${\mathbf S}_{D} = \big[{\mathbf
s}_{1}~{\mathbf s}_{2}~\ldots~ {\mathbf s}_{D} \big]$, in which the
basis vectors ${\mathbf s}_{d}$, $d=1,2, \ldots, D$ will be obtained
via an optimization algorithm and iterations between the ${\mathbf
S}_{D}$ and ${\mathbf w}_D$ will be performed. The JIO method
consists of solving the following optimization problem
\begin{equation}
\begin{split}
\big[ {\boldsymbol S}_{D,{\rm opt}}, {\boldsymbol w}_{D,{\rm opt}}
\big] & = \arg \min_{ {\boldsymbol S}_{D}, {\boldsymbol w}_D}
\underbrace{ \underbrace{{\boldsymbol w}^{H}_D {\boldsymbol
S}_{D}^H{\boldsymbol R}{\boldsymbol S}_{D}{\boldsymbol w}_D
}_{x(i)}}_{{\boldsymbol C} ({\boldsymbol S}_{D}, {\boldsymbol
w}_{D})},\\ & {\rm subject} ~ {\rm to}~ {\boldsymbol
w}_{D}^H{\boldsymbol S}_{D}^H {\boldsymbol s} = 1 \label{lsjio}
\end{split}
\end{equation}
where it should be remarked that the optimization problem in
(\ref{lsjio}) is non convex, however, the algorithms do not present
convergence problems. Numerical studies with JIO methods indicate
that the minima are identical and global. Proofs of global
convergence have been established with different versions of JIO
schemes \cite{delamare_spl07,fa,delamareel,delamare_esp,fa2011} ,
which demonstrate that a least-squares (LS) algorithm converges to
the reduced-rank Wiener filter.

In order to solve the above problem, we resort to the method of
Lagrange multipliers \cite{haykin} and transform the constrained
optimization into an unconstrained one expressed by the Lagrangian
\begin{equation}
\begin{split}
{\mathcal L}({\boldsymbol S}_D, {\boldsymbol w}_D) & = {\boldsymbol
w}^{H}_D {\boldsymbol S}_{D}^H{\boldsymbol R}{\boldsymbol
S}_{D}{\boldsymbol w}_D  + \lambda (\bar{\boldsymbol w}^H_D{\bf
S}_D^H{\boldsymbol s}-1) , \label{uopt}
\end{split}
\end{equation}
where $\lambda$ is a scalar Lagrange multiplier. By fixing
${\boldsymbol w}_D$, minimizing (\ref{uopt}) with respect to
${\boldsymbol S}_D$ and solving for $\lambda$, we obtain
\begin{equation}
\begin{split}
{\boldsymbol S}_D & = \frac{ {\boldsymbol R}^{-1} {\boldsymbol s}
{\boldsymbol w}^H_D {\boldsymbol R}_{\bar{w}}^{-1}} {{\boldsymbol
w}^H_D {\boldsymbol R}_{\bar{w}}^{-1} {\boldsymbol w}_D {\boldsymbol
s}^H {\boldsymbol R}^{-1} {\boldsymbol s}}, \label{filts}
\end{split}
\end{equation}
where  ${\boldsymbol R} = E[{\boldsymbol r}(i){\boldsymbol
r}^{H}(i)]$ and ${\boldsymbol R}_{\bar{w}} = E[\bar{\boldsymbol
w}_D\bar{\boldsymbol w}^{H}_D]$. By fixing ${\boldsymbol S}_D$,
minimizing (\ref{uopt}) with respect to ${\boldsymbol w}_D$ and
solving for $\lambda$, we arrive at the expression
\begin{equation}
{\boldsymbol w}_D =  \frac{ \bar{\boldsymbol R}^{-1} {\boldsymbol
s}}{{\boldsymbol s}^H {\boldsymbol R}^{-1} {\boldsymbol s}},
\label{filtw}
\end{equation}
where ${\boldsymbol R}_D = E[{\boldsymbol S}_D^H{\boldsymbol
r}(i){\boldsymbol r}^H(i)  {\boldsymbol S}_D] =E[{\boldsymbol
r}_D(i) {\boldsymbol r}^{H}_D(i)]$, ${\boldsymbol s}_D =
{\boldsymbol S}_D^H{\boldsymbol s}$. Note that the expressions in
(\ref{filts}) and (\ref{filtw}) are not closed-form solutions for
${\boldsymbol w}_D$ and ${\boldsymbol S}_D$ since (\ref{filts}) is a
function of ${\boldsymbol w}_D$ and (\ref{filtw}) depends on
${\boldsymbol S}_D$. Thus, it is necessary to iterate (\ref{filts})
and (\ref{filtw}) with initial values to obtain a solution. Unlike
the Krylov subspace-based methods \cite{goldstein} and the AVF
\cite{pados01} methods, the JIO scheme provides an iterative
exchange of information between the low-rank beamformer and the
rank-reduction matrix and leads to a simpler adaptive
implementation. The key strategy lies in the joint optimization of
the filters. The rank $D$ must be set by the designer to ensure
appropriate performance or can be estimated via another algorithm.
In terms of complexity, the JIO techniques have a computational cost
that is related to the optimization algorithm. With recursive LS
algorithms the complexity is quadratic with $M$ ($(O(M^2)$), whereas
the complexity can be as low as linear with $M$ when stochastic
gradient algorithms are adopted \cite{jidf}.

\subsection{JIDF-based algorithms}

This section reviews a low-rank space-time beamforming technique
based on the joint interpolation, decimation and filtering (JIDF)
concept \cite{delamare_icassp07,jidf,RuiSTAP2010}. The JIDF approach
allows a designer to compute the parameters of the rank-reduction
matrix and the low-rank space-time beamformer with a low complexity.
The motivation for designing a rank-reduction matrix based on
interpolation and decimation comes from two observations. The first
is that rank reduction can be performed by constructing new samples
with interpolators and eliminating (decimating) samples that are not
useful in the STAP design. The second comes from the structure of
the rank-reduction matrix, whose columns are a set of vectors formed
by the interpolators and the decimators.

In the JIDF scheme, the number of elements for adaptive processing
is substantially reduced, resulting in considerable computational
savings and very fast convergence performance for the radar
applications. The $M \times 1$ received vector ${\boldsymbol r}(i)$
is processed by a multiple processing branch (MPB) scheme with $B$
branches, where each spatio-temporal processing branch contains an
interpolator, a decimation unit and a low-rank space-time
beamformer. In the $b$-th branch, the received vector ${\boldsymbol
r}(i)$ is filtered by the interpolator ${\boldsymbol v}_b =
[v_{b,0}~v_{b,1}~\ldots~ v_{b,I-1}]^T$ with $I$ coefficients,
resulting in an interpolated received vector ${\boldsymbol r}_b(i)$
with $M$ samples, which is expressed by
\begin{equation}
{\boldsymbol r}_b(i) = {\boldsymbol V}_b^H{\boldsymbol r}(i),
\end{equation}
where the $M \times M$ Toeplitz convolution matrix is given by
\begin{equation}
{\boldsymbol V}_b = \left[ \begin{array}{cccc} v_{b,0} & 0 & \ldots & 0\\
\vdots & v_{b,0} & \vdots & \vdots \\ v_{b,I-1} & \vdots & \vdots & 0 \\
0 & v_{b,I-1} & \vdots &  0\\ 0 & 0 & \vdots & \vdots \\ \vdots &
\vdots & \vdots & 0 \\ 0 & 0 & \vdots & v_{b,0}
\end{array} \right]
\end{equation}
The vector ${\boldsymbol r}_b(i)$ can be expressed in an alternative
way that is useful for the design of the JIDF scheme and is
described by
\begin{equation}
{\boldsymbol r}_b(i) = {\boldsymbol V}_b^H{\boldsymbol r}(i) =
{\boldsymbol \Re}_0(i){\boldsymbol v}_b,
\end{equation}
where the $M \times I$ matrix ${\boldsymbol \Re}_o(i)$ with the
samples of ${\boldsymbol r}(i)$ has a Hankel structure and is
described by
\begin{equation}
{\boldsymbol \Re}_o(i) = \left[ \begin{array}{cccc} r_{0}(i) & r_{1}(i)   & \ldots & r_{{\rm I}-1}(i)  \\
r_{1}(i)  & r_{2}(i)   & \ldots & r_{{\rm I}}(i)  \\
\vdots & \vdots  & \ddots & \vdots \\
r_{M-2}(i)  & r_{M-1}(i)  & \ldots & 0  \\
r_{M-1}(i)  & 0  & \ldots & 0  \\ \end{array} \right]
\end{equation}
The dimensionality reduction is performed by a decimation unit with
$D \times M$ decimation matrices ${\boldsymbol D}_b$ that transforms
${\boldsymbol r}_I(i)$ into $D\times 1$ vectors ${\boldsymbol
r}_{D,b}(i)$ with $b = 1, \ldots, B$, where $D = M/L$ is the rank of
the resulting system of equations that will be generated and $L$ is
the decimation factor. The $D \times 1$ vector ${\boldsymbol
r}_{D,b}(i)$ for branch $b$ is expressed by
\begin{equation}
\begin{split}
{\boldsymbol r}_{D,b} & = {\boldsymbol S}_{D,b}^H {\boldsymbol r}(i)
= {\boldsymbol D}_{D,b}{\boldsymbol V}_b^H{\boldsymbol r}(i) \\
& = {\boldsymbol D}_{D,b} {\boldsymbol \Re}_o(i) {\boldsymbol v}
\end{split}
\end{equation}
where ${\boldsymbol S}_{D,b}$ is the rank-reduction matrix and the
vector ${\boldsymbol r}_{D,b}(i)$ for branch $b$ is used in the
minimization of the output power for branch $b$. The output at the
end of the JIDF scheme $y(i)$ is selected according to
\begin{equation}
y(i) = y_{b_s}(i) ~~~ {\rm when}~~~ b_s = \arg \min_b |y_b|^2
\end{equation}
where $B$ is a parameter to be set by the designer. For the
computation of the parameters of the JIDF scheme, it is fundamental
to express the output $y_b(i)$ as a function of the interpolator
${\boldsymbol v}_b$, the decimation matrix ${\boldsymbol D}_{D,b}$
and the low-rank space-time beamformer ${\boldsymbol w}_{D,b}$ as
follows:
\begin{equation}
\begin{split}
y_b(i) & = {\boldsymbol w}^H_{D,b} {\boldsymbol S}_{D,b}^H
{\boldsymbol r}(i) \\ & = {\boldsymbol w}^H_{D,b} {\boldsymbol
D}_{D,b} {\boldsymbol \Re}_o(i) {\boldsymbol v}, \label{yb_out}
\end{split}
\end{equation}
where the expression (\ref{yb_out}) indicates that the
dimensionality reduction carried out by the JIDF scheme depends on
finding appropriate ${\boldsymbol v}_b$, ${\boldsymbol D}_{D,b}$ and
${\boldsymbol w}_{D,b}$. Unlike the previously discussed low-rank
beamforming techniques, the JIDF is able to substantially reduce the
cost of the rank-reduction matrix.

The parameters of the JIDF scheme that perform low-rank space-time
MVDR beamforming can be computed by solving the following
optimization problem
\begin{equation}
\begin{split}
[{\boldsymbol w}_{D,{\rm opt}}, {\boldsymbol v}_{\rm opt},
{\boldsymbol D}_{D, b_s}] & = \arg \min_{{\boldsymbol w}_{D,b},
{\boldsymbol v}_b, {\boldsymbol D}_{D,b}} {\boldsymbol w}^{H}_{D,b}
E[{\boldsymbol D}_{D,b}{\boldsymbol \Re}_o(i) {\boldsymbol v}
{\boldsymbol v}^H {\boldsymbol \Re}_o^H(i) {\boldsymbol D}_{D,b}^H]
{\boldsymbol
w}_{D}\\
& {\rm subject}~{\rm to}~ {\boldsymbol w}_{D,b}^H{\boldsymbol
D}_{D,b}{\boldsymbol S}_o {\boldsymbol v}_b = 1, \label{jidf_opt}
\end{split}
\end{equation}
where ${\boldsymbol S}_o$ is $M \times I$ steering matrix with a
Hankel structure, which has the same form as ${\boldsymbol
\Re}_o(i)$ and is given by
\begin{equation}
{\boldsymbol S}_o(i) = \left[ \begin{array}{cccc} s_{0}(i) & s_{1}(i)   & \ldots & s_{{\rm I}-1}(i)  \\
s_{1}(i)  & s_{2}(i)   & \ldots & s_{{\rm I}}(i)  \\
\vdots & \vdots  & \ddots & \vdots \\
s_{M-2}(i)  & s_{M-1}(i)  & \ldots & 0  \\
s_{M-1}(i)  & 0  & \ldots & 0  \\ \end{array} \right].
\end{equation}
The constrained optimization in (\ref{jidf_opt}) can be transformed
into an unconstrained optimization problem by using the method of
Lagrange multipliers, which results in
\begin{equation}
{\mathcal L}({\boldsymbol w}_{D,b}, {\boldsymbol v}_b, {\boldsymbol
D}_{D,b}) = {\boldsymbol w}^{H}_{D}  E[{\boldsymbol
D}_{D,b}{\boldsymbol \Re}_o(i) {\boldsymbol v} {\boldsymbol v}^H
{\boldsymbol \Re}_o^H(i) {\boldsymbol D}_{D,b}^H] {\boldsymbol
w}_{D} + \lambda({\boldsymbol w}_{D}^H{\boldsymbol
D}_{D,b}{\boldsymbol S}_o {\boldsymbol v} - 1), \label{Lag_jidf}
\end{equation}
where $\lambda$ is a Lagrange multiplier.

The strategy to compute the parameters of the low-rank space-time
beamformer based on the JIDF scheme is to minimize the cost function
with respect to a set of parameters and fix the remaining
parameters. By minimizing (\ref{Lag_jidf}) with respect to
${\boldsymbol v}_b$, we obtain
\begin{equation}
{\boldsymbol v}_b = \frac{{\boldsymbol R}_{v,b}^{-1} {\boldsymbol
s}_{v,b}}{{\boldsymbol s}_{v,b}^H {\boldsymbol
R}_{v,b}^{-1}{\boldsymbol s}_{v,b}}, \label{v_des}
\end{equation}
where ${\boldsymbol R}_{v,b} = E[{\boldsymbol r}_{v,b}{\boldsymbol
r}_{v,b}^H]$ is the $I \times I$ autocorrelation matrix,
${\boldsymbol r}_{v,b} = {\boldsymbol D}_{D,b}^H{\boldsymbol
R}_o^H{\boldsymbol w}_{D,b}$, and ${\boldsymbol s}_{v,b} =
{\boldsymbol D}_{D,b}^H{\boldsymbol S}_o^H{\boldsymbol w}_{D,b}$ is
the $I\times$ low-rank steering vector. By minimizing
(\ref{Lag_jidf}) with respect to ${\boldsymbol w}_{D,b}$, we have
\begin{equation}
{\boldsymbol w}_{D,b} = \frac{{\boldsymbol R}_{w,b}^{-1}
{\boldsymbol s}_{w,b}}{{\boldsymbol s}_{w,b}^H {\boldsymbol
R}_{w,b}^{-1}{\boldsymbol s}_{w,b}}, \label{w_des}
\end{equation}
where ${\boldsymbol R}_{w,b} = E[{\boldsymbol r}_{w,b}{\boldsymbol
r}_{w,b}^H]$ is the $D \times D$ autocorrelation matrix,
${\boldsymbol r}_{w,b} = {\boldsymbol D}_{D,b}{\boldsymbol
R}_o{\boldsymbol v}_{b}$, and ${\boldsymbol s}_{w,b} = {\boldsymbol
D}_{D,b}{\boldsymbol S}_o{\boldsymbol v}_{b}$ is the $D\times I$
low-rank steering vector. In order to compute ${\boldsymbol v}_{b}$
and ${\boldsymbol w}_{D,b}$, a designer needs to iterate them for
each processing branch $b$.

The decimation matrix ${\boldsymbol D}_{D,b}$ is selected to
minimize the square of the output of the beamformer $y_b(i)$
obtained for all the $B$ branches
\begin{equation}
{\boldsymbol D}_{D,b} = {\boldsymbol D}_{D,b_{\rm s}}[i] ~~
\textrm{when} ~~ b_{\rm s} = \arg \min_{1\leq b \leq B}
|y_{b}(i)|^{2}, \label{D_sel}
\end{equation}
The design of the decimation matrix ${\boldsymbol D}_{D,b}$ imposes
constraints on the values of the elements of the matrix such that
they only take the value zero or one. Since the optimal approach for
the design of ${\boldsymbol D}_{D,b}$ corresponds to an exhaustive
search, we consider a suboptimal technique that employs pre-stored
patterns. The decimation scheme employs a structure formed in the
following way
\begin{equation}
{\boldsymbol S}_{D,b} = [ {\boldsymbol \phi}_{b,1}~{\boldsymbol
\phi}_{b,2}~{\boldsymbol \phi}_{b,D}],
\end{equation}
where ${\boldsymbol \phi}_{b,d}$ is an $M \times 1$ vector composed
of a single one and zeros elsewhere as described by
\begin{equation}
{\boldsymbol \phi}_{b,d} = [ \underbrace{0,~\ldots,~ 0}_{z_{b,d}},
~1, ~\underbrace{0,~\ldots,~ 0}_{M-z_{b,d}-1}],
\end{equation}
where $z_{b,d}$ is the number of zeros before the only element equal
to one. We set the value of $z_{b,d}$ in a deterministic way which
can be expressed as
\begin{equation}
z_{b,d} = \frac{M}{D} \times (d - 1) + (b - 1).
\end{equation}
It is necessary to iterate (\ref{v_des}), (\ref{w_des}) and
(\ref{D_sel}) in an alternated form (one followed by the other) with
an initial value to obtain a solution. The expectations can be
estimated either via time averages or by instantaneous estimates and
with the help of adaptive algorithms.

\section{Sparsity-Aware Beamforming Algorithms} \label{sec4}

This section considers space-time beamforming algorithms that
exploit the sparsity encountered in the data processed by radar
systems. In particular, the motivation for exploiting the sparsity
of data vectors observed by radar systems is given and a brief
discussion on the suitability of sparsity-aware algorithms for radar
applications is provided. A general approach to design space-time
beamforming algorithms based on the $l_1$-norm regularization is
described. The main principle is to employ a reduced number of
weights to suppress the clutter and the jamming signals encountered
in radar applications.

Recently, motivated by compressive sensing (CS) techniques used in
radar, several authors have considered CS ideas for moving target
indication (MTI) and STAP problems \cite{maria}-\cite{selesnick}.
The core notion in CS is to regularize a linear inverse problem by
including prior knowledge that the signal of interest is sparse
\cite{parker}. These works on space-time beamforming techniques
based on CS rely on the recovery of the clutter power in
angle-Doppler plane, which is usually carried out via convex
optimization tools. However, these methods are based on linear
programming and have a quite high computational complexity
($O(K^3)$), where $K$ is the dimension of the angle-Doppler plane.
In this section, we describe the concept of a sparsity-aware STAP
(SA-STAP) algorithm that can improve the detection capability using
a small number of snapshot. To overcome the high complexity of the
CS-STAP type algorithm, we design the STAP algorithm with another
strategy, by imposing the sparse regularization to the minimum
variance (MV) cost function. Since the interference variance has
often a low-rank property, we assume that a number of samples of the
data cube are not meaningful for processing and the optimal STAP
beamformer is sparse, or nearly sparse. Then, we exploit this
feature by using a $l_1$-norm regularization. With this motivation,
the STAP algorithm design becomes a mixed $l_1$-norm and $l_2$-norm
optimization problem.

The conventional space-time beamforming algorithms do not exploit
the sparsity of the received signals. In this exposition, it is
assumed that a number of samples of the data cube are not meaningful
for processing and a reduced number of active weights of the
space-time beamformer can effectively suppress the clutter and the
jamming signals. Specifically, a sparse regularization is imposed to
the space-time MVDR beamforming design. Thus, the space-time
beamformer design can be described as the following optimization
problem
\begin{equation}
\begin{split}
{\boldsymbol w}_{\rm opt} & = \arg \min_{{\boldsymbol w}}
{\boldsymbol w}^H {\boldsymbol R}{\boldsymbol w}\\
{\rm subject}~{\rm to}~ & {\boldsymbol w}^{H}{\boldsymbol s} =
1~{\rm and}~||{\boldsymbol w}||_1 = 0,
\end{split}
\end{equation}
where the objective of the $l_1$-norm regularization is to force the
components of the space-time beamformer ${\boldsymbol w}$ to zero
\cite{angelosante}. This problem can be solved using the method of
Lagrange multipliers, which results in the following unconstrained
cost function
\begin{equation}
\begin{split}
{\mathcal L}({\boldsymbol w},\alpha,\lambda) & = {\boldsymbol w}^H
{\boldsymbol R}{\boldsymbol w} +  \alpha({\boldsymbol
w}^{H}{\boldsymbol s} - 1) + \lambda (||{\boldsymbol w}||_1),
\end{split}
\end{equation}
The unconstrained cost function above is convex, however, it is
non-differentiable which makes it difficult for one to use the
method of Lagrange Multipliers directly and obtain an expression for
the space-time beamformer. To this end, the following approximation
to the regularization term is employed
\begin{equation}
||{\boldsymbol w}||_1 \approx  {\boldsymbol w}^H {\boldsymbol
\Lambda} {\boldsymbol w},
\end{equation}
where
\begin{equation}
{\boldsymbol \Lambda} = {\rm diag}  \left(\frac{1}{|w_1|+ \epsilon }
~ \frac{1}{|w_2| + \epsilon} ~ \ldots ~ \frac{1}{|w_M|+ \epsilon}
\right) ,
\end{equation}
where $\epsilon$ is a small positive constant. Simultaneously, we
assume that the partial derivative of ${\boldsymbol w}^H
{\boldsymbol \Lambda} {\boldsymbol w}$ with respect to ${\boldsymbol
w}^*$ is given by
\begin{equation}
\frac{ \partial {\boldsymbol w}^H {\boldsymbol \Lambda }
{\boldsymbol w} } {\partial {\boldsymbol w}^*} \approx {\boldsymbol
\Lambda} {\boldsymbol w}.
\end{equation}
With the development above, an approximation to the unconstrained
cost function can be employed as described by
\begin{equation}
\begin{split}
{\mathcal L}({\boldsymbol w},\alpha,\lambda) & \approx {\boldsymbol
w}^H {\boldsymbol R}{\boldsymbol w} +  \alpha({\boldsymbol
w}^{H}{\boldsymbol s} - 1) + \lambda {\boldsymbol w}^H {\boldsymbol
\Lambda} {\boldsymbol w},
\end{split}
\end{equation}
By computing the gradient terms with respect to ${\boldsymbol w}^*$
and $\alpha$ and equating them to zero, we obtain the following
expression for the space-time beamformer
\begin{equation}
{\boldsymbol w} = \frac{({\boldsymbol R} + \lambda {\boldsymbol
\Lambda} )^{-1}{\boldsymbol s}}{{\boldsymbol s}^H ({\boldsymbol R} +
\lambda {\boldsymbol \Lambda} )^{-1}{\boldsymbol s}}.
\label{sparse_bf}
\end{equation}
Comparing (\ref{sparse_bf}) with the conventional optimal space-time
beamformer in (\ref{conv_bf}), we find that there is an additional
term $\lambda{\boldsymbol \Lambda}$ in the inverse of the
interference covariance matrix ${\boldsymbol R}$, which is due to
the $l_1$-norm regularization. The term $\lambda$ is a positive
scalar which provides a trade-off between the sparsity and the
output interference power. The larger the chosen $\lambda$, the more
components are shrunk to zero \cite{zibulevsky}. It should also be
remarked that the expression for the beamformer in (\ref{sparse_bf})
is not a closed-form solution since ${\boldsymbol \Lambda}$ is a
function of ${\boldsymbol w}$. Thus it is necessary to develop an
iterative procedure to compute the parameters of the space-time
beamformer.

\section{Knowledge-Aided Beamforming Algorithms} \label{sec5}

Although STAP techniques are considered efficient tools for
detection of slow targets by airborne radar systems in strong
clutter environments \cite{klemm}, due to the very large number of
degrees of freedom (DoFs) conventional space-time beamformers have a
slow convergence and require about twice the DoFs of the independent
and identically distributed (IID) training snapshots to yield an
average performance loss of roughly $3$dB \cite{ward}. In real
scenarios, it is hard to obtain so many IID training snapshots,
especially in heterogeneous environments. Low-rank
\cite{guerci}-\cite{jidf} and sparsity-aware
\cite{maria}-\cite{zhaocheng} methods have been considered to
counteract the slow convergence of the conventional space-time
beamformers. Nevertheless, there are other alternatives to improve
the training of STAP algorithms and improve their performance. These
other methods can also be combined with the techniques previously
discussed. Recently developed knowledge-aided (KA) STAP algorithms
have received a growing interest and become a key concept for the
next generation of adaptive radar systems
\cite{wicks06}-\cite{fa10b}. The core idea of KA-STAP is to
incorporate prior knowledge, provided by digital elevation maps,
land cover databases, road maps, the Global Positioning System
(GPS), previous scanning data and other known features, to compute
estimates of the clutter covariance matrix with high accuracy
\cite{melvin06a}. Prior work on KA-STAP algorithms include the
exploitation of prior knowledge of the clutter ridge to form the
STAP filter weights \cite{melvin06b}, use of prior knowledge about
the terrain \cite{capraro} and prior knowledge about the covariance
matrix of the clutter and the jamming signals
\cite{blunt}-\cite{fa10b}.

In this section, we discuss a strategy to mitigate the deleterious
effects of the heterogeneity in the secondary data, which makes use
of a priori knowledge of the clutter covariance matrix and has
recently gained significant attention in the literature
\cite{wicks06}-\cite{fa10b}. In KA-STAP techniques, there are two
basic tasks that need to be addressed: the first one is how to
obtain prior knowledge from the terrain knowledge of the clutter and
how to estimate the real interference covariance matrix with the
prior knowledge \cite{wicks06}-\cite{capraro} and the second is how
to apply the covariance matrix estimates in the design of the
space-time beamforming algorithm \cite{blunt}-\cite{fa10b}. We first
review how a designer can obtain prior knowledge of the clutter and
employ this knowledge to build a known covariance matrix
${\boldsymbol R}_o$. Then, we present a method to combine this prior
knowledge with commonly used estimation techniques to compute the
covariance matrix of the received vector ${\boldsymbol r}(i)$,
resulting in a combined covariance matrix estimate $\hat{\boldsymbol
R}_c$ for use in the space-time beamformer that is more accurate and
has an enhanced performance.

The optimal space-time beamformer employs the following expression
to compute its parameters
\begin{equation}
{\boldsymbol w} = \frac{\hat{\boldsymbol R}^{-1}{\boldsymbol
s}}{{\boldsymbol s}^H \hat{\boldsymbol R}^{-1}{\boldsymbol s}},
\label{conv_bf_est}
\end{equation}
where an estimate of the covariance matrix is typically obtained by
\begin{equation}
\hat{\boldsymbol R} = \frac{1}{K} \sum_{k=1}^{K} {\boldsymbol
r}(k){\boldsymbol r}^H(k),
\end{equation}
where  ${\boldsymbol r}(k)$ is taken from secondary data. The
estimate $\hat{\boldsymbol R}$ can be sufficiently accurate when $K$
is at least twice as great as $M$ \cite{brennan} and the training
samples are assumed i.i.d. However, it is by now well understood
that the clutter environments are often heterogeneous and this leads
to performance degradation on space-time beamforming. KA-STAP
techniques can significantly help to combat the heterogeneity
\cite{stoica}.

With KA techniques the covariance matrix ${\boldsymbol R}_c$ is
estimated by combining an initial guess of the covariance matrix
${\boldsymbol R}_o$ derived from the digital terrain database or the
data probed by radar in previous scans, and the sample average
covariance matrix estimate in the present scan $\hat{\boldsymbol R}$
so that
\begin{equation}
{\boldsymbol R}_c = \alpha {\boldsymbol R}_o + (1-\alpha)
\hat{\boldsymbol R},
\end{equation}
where $0\leq \alpha \leq 1$. Alternatively, this principle can be
applied to the inverse of the covariance matrix estimate
\begin{equation}
{\boldsymbol R}_c^{-1} = \eta {\boldsymbol R}_o^{-1} + (1-\eta)
\hat{\boldsymbol R}^{-1},
\end{equation}
where $0\leq \eta \leq 1$.

In order to compute the parameter $\eta$, we need to consider the
optimization problem
\begin{equation}
\eta_{\rm opt} = \arg \min_{\eta} {\boldsymbol w}^H {\boldsymbol
R}{\boldsymbol w}, \label{eta_opt}
\end{equation}
where we use the relation
\begin{equation}
{\boldsymbol w}(i) = \eta {\boldsymbol w}_{o} + (1-\eta)
\hat{\boldsymbol w}(i),
\end{equation}
where ${\boldsymbol w}(i) = {\boldsymbol R}^{-1} {\boldsymbol s}$,
${\boldsymbol w}_o = {\boldsymbol R}^{-1}_o {\boldsymbol s}$ and
$\hat{\boldsymbol w} = \hat{\boldsymbol R}^{-1} {\boldsymbol s}$. We
can obtain the optimal value for $\eta$ by equating the gradient of
the cost function in (\ref{eta_opt}) to zero, which results in
\cite{stoica}
\begin{equation}
\eta_{\rm opt} = \frac{ \Re ({\boldsymbol s}^H( \hat{\boldsymbol
R}^{-1} - {\boldsymbol R}_o^{-1} ) {\boldsymbol R} \hat{\boldsymbol
R}^{-1} {\boldsymbol s}) }{ {\boldsymbol s}^H({\boldsymbol R}_o^{-1}
- \hat{\boldsymbol R}^{-1} ) {\boldsymbol R} ( {\boldsymbol
R}_o^{-1} - \hat{\boldsymbol R}^{-1} ){\boldsymbol s} }
\end{equation}
Since ${\boldsymbol R}$ above is unknown, we have to estimate it in
real time using either time averages or adaptive algorithms.

\section{Simulations} \label{sec6}

In this section, we assess the performance of the space-time
beamforming algorithms discussed in this chapter using simulated
radar data. Specifically, we consider the optimal MVDR space-time
beamforming algorithm that assumes perfect knowledge of the
covariance matrix of the received data, and the MVDR space-time
beamformer using the sample matrix inversion (SMI-MVDR). The
low-rank space-time beamforming algorithms using EVD (LR-EVD), the
Krylov subspace approach (LR-Krylov), the JIO (LR-JIO) and the JIDF
(LR-JIDF) algorithms are also considered with a rank equal to $D$.
We also consider the sparsity-aware (SA-MVDR) and the
knowledged-aided (KA-MVDR) space-time beamforming algorithms. All
the analyzed algorithms estimate the statistical quantities via
time-averages in a similar way to a least-squares method. The
parameters of the simulated radar platform are shown in Table
\ref{tabpar}. For all simulations, we assume the presence of a
mixture of two broadband jammers at $-45^o$ and $60^o$ with
jammer-to-noise ratio (JNR) equal to $40$ dB. The
clutter-to-noise-ratio (CNR) is fixed at $40$ dB. All the results
presented are averages over 1000 independent Monte-Carlo runs.

\begin{table}[h]
\centering \caption{Airborne radar system
parameters.}
\begin{tabular}{|l|c|} \hline\hline Parameter & Value\\
[0.5ex] \hline
Antenna array ~~~~~~~~~~~~~~~& Sideway-looking array (SLA) \\
Carrier frequency ($f_c$)~~~~~~~ & ~~~~~~~450 MHz \\
Transmit pattern & ~~~~~~~~Uniform \\
PRF ($f_r$)~~~~~~~  & ~~~~~~~~300 Hz \\
Platform velocity ($v$)~~~~~~~ & ~~~~~~~75 m/s \\
Platform height ($h$)~~~~~~~ & ~~~~~~~9000 m \\
Clutter-to-Noise ratio (CNR)~~~~~~~ & ~~~~~~~40 dB \\
Elements of sensors ($N$)~~~~~~~ & ~~~~~~~8\\
Number of Pulses ($J$)~~~~~~~ & ~~~~~~~8 \\[1ex]
\hline
\end{tabular}
\label{tabpar}
\end{table}

\begin{figure}[h]
\begin{center}
\def\epsfsize#1#2{1\columnwidth}
\epsfbox{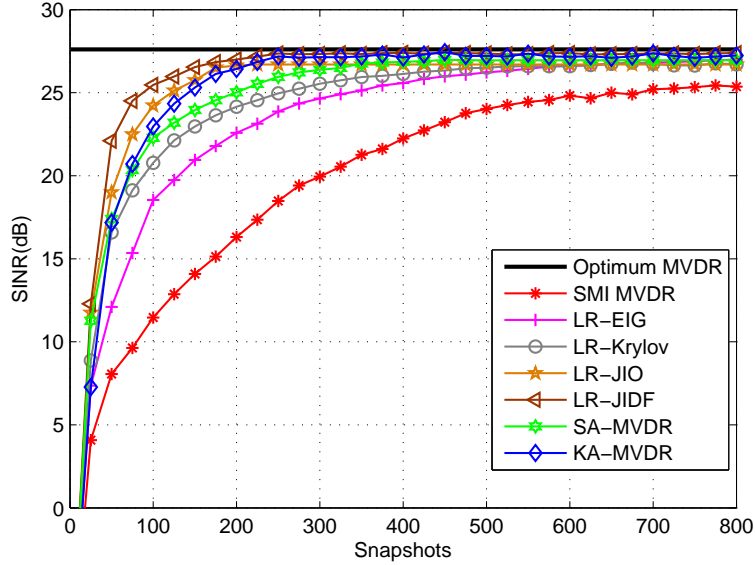} \vspace*{-1em} \caption{SINR Performance against
the snapshots. Parameters: $B=8$, $I=8$ and $D=6$, $M = 64$,
$\bar{\boldsymbol R}[0] = 0.01 {\boldsymbol I}$, $\epsilon =
0.1$.}\label{sinrvs}
\end{center}
\end{figure}

In the first experiment, we assess the SINR performance of the
different space-time beamforming algorithms as shown in Fig.
\ref{sinrvs}. The algorithms are simulated over $800$ snapshots and
the SNR is set to $10$ dB. The results show that the LR-JIDF
algorithm achieves the best results, followed by the LR-JIO, the
KA-MVDR, the SA-MVDR, the LR-Krylov, the LR-EIG and the SMI MVDR
algorithms. The curves indicate that the use of low-rank algorithms
is highly beneficial to the performance of space-time beamforming
algorithms in radar systems. In particular, the LR-JIDF and LR-JIO
algorithms have a very fast convergence performance. It should also
be remarked that the SA-MVDR and KA-MVDR algorithms obtain a
performance that is significantly better than the conventional SMI
MVDR algorithm. Since the SA-MVDR and KA-MVDR techniques are
modifications of the SMI MVDR techniques exploiting sparsity and
prior knowledge about the covariance matrix, respectively, it is
interesting to note that by exploiting these properties it is
possible to significantly improve the performance of beamforming
algorithms.

\begin{figure}[h]
\begin{center}
\def\epsfsize#1#2{1\columnwidth}
\epsfbox{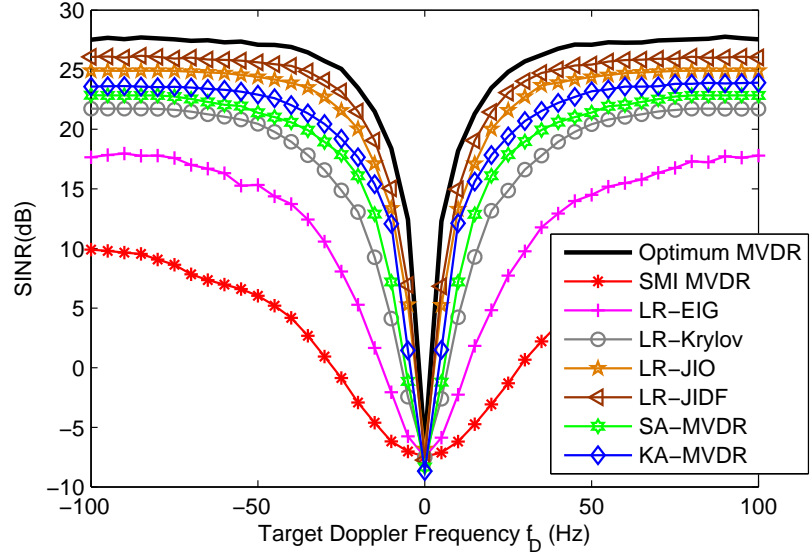} \vspace*{-1em} \caption{SINR Performance against
the target Doppler frequency. Parameters: $B=8$, $I=8$ and $D=6$, $M
= 64$, $\bar{\boldsymbol R}[0] = 0.01 {\boldsymbol I}$, $\epsilon =
0.1$.}\label{sinrvfd}
\end{center}
\end{figure}

We evaluate the SINR performance against the target Doppler
frequency at the main bean look angle for our proposed algorithms
and other existing algorithms, which are illustrated in Fig.
\ref{sinrvfd}. The potential Doppler frequency space form $-100$ to
$100$ Hz is examined and 100 snapshots are used to train the
beamformers. The plots show that the analyzed algorithms converge
and approach the optimum in a short time, and form a deep null to
cancel the main beam clutter. Again, the LR-JIDF algorithm
outperforms the other analyzed algorithms.

\begin{figure}[h]
\begin{center}
\def\epsfsize#1#2{1\columnwidth}
\epsfbox{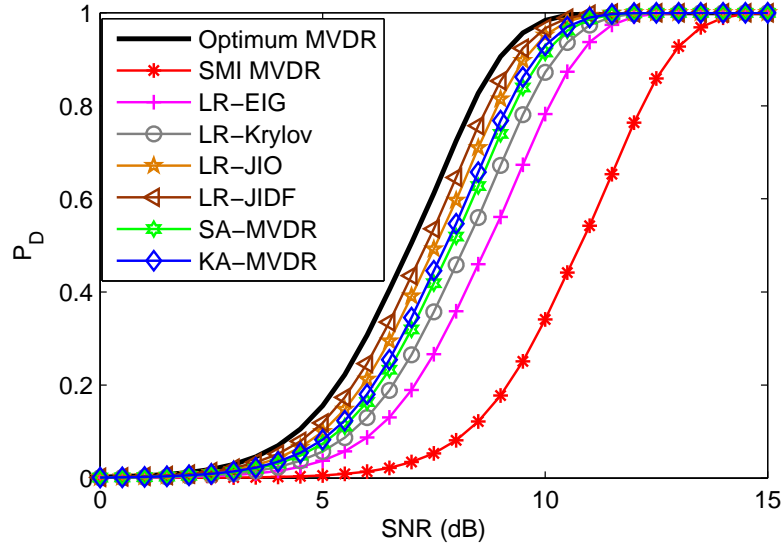} \vspace*{-1em} \caption{Probability of detection
against SNR.Parameters: $B=8$, $I=8$ and $D=6$, $M = 64$,
$\bar{\boldsymbol R}[0] = 0.01 {\boldsymbol I}$, $\epsilon =
0.1$.}\label{Pdvsnr}
\end{center}
\end{figure}

In the third example, the probability of detection $P_D$ versus SNR
performance is presented for all schemes using $200$ snapshots as
the training data as shown in Fig. \ref{Pdvsnr}. The false alarm
rate PFA is set to $10^{-6}$ and we suppose the target injected in
the boresight ($0^0$) with Doppler frequency $100$Hz. The figure
illustrates that the analyzed algorithms provide sub-optimal
detection performance using short support data. Note that for $P_D =
0.9$ ($90 \%$ percent), the LR-JIDF  and LR-JIO schemes are within
less than $1$ dB from the performance of the optimal MVDR algorithm.
The remaining techniques exhibit increasing performance losses as
compared to the optimal MVDR algorithm and it should be noted that
the conventional SMI MVDR method has a performance degradation of up
to $5$ dB for the same performance measured in terms of $P_D$. This
suggest that the application of more sophistical space-time
beamforming algorithms is key to achieving an improved performance.

\begin{figure}[!htb]
\begin{center}
\def\epsfsize#1#2{1\columnwidth}
\epsfbox{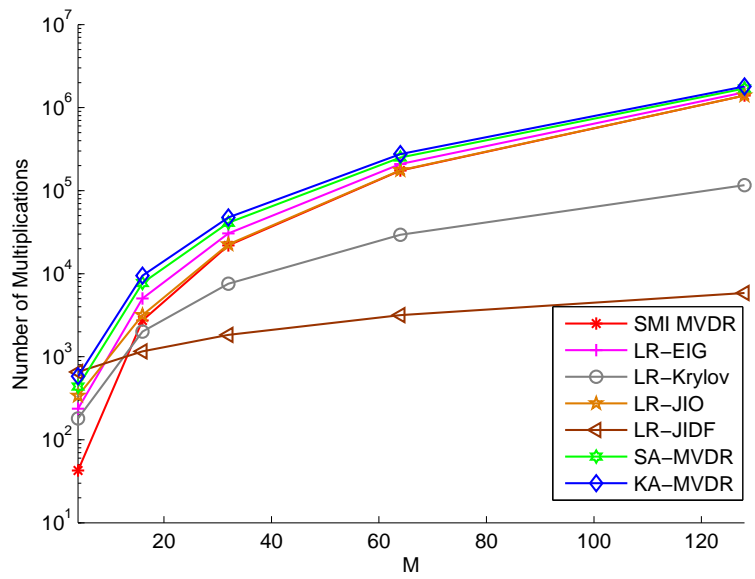} \vspace*{-1em} \caption{Computational complexity
in terms of multiplications of the analyzed space-time beamforming
algorithms. Parameters: $B=8$, $I=8$ and $D=6$.} \label{comp}
\end{center}
\end{figure}

At last, we detail the computational complexity in terms of
multiplications of the analyzed schemes in Fig. \ref{comp}. The
curves show that the computational complexity of the LR-JIDF and
LR-Krylov algorithms is significantly lower than the remaining
algorithms. Indeed, there is a significant computational advantage
obtained by using the LR-JIDF and LR-Krylov algorithms and this
advantage becomes more pronounced as $M$ is increased. The other
analyzed algorithms have a complexity that scales as a cubic
function of $M$. This high complexity can be mitigated by the use of
adaptive algorithms, which can reduce the computational cost by at
least an order of magnitude.

\section{Concluding Remarks} \label{sec7}

This chapter has presented recent advances on space-time beamforming
algorithms for phased-array radar systems and investigated their
performance via computer simulations. Specifically, we have
considered in detail some of the most successful space-time
beamforming algorithms that exploit low-rank and sparsity properties
as well as the use of prior-knowledge to improve the performance of
STAP algorithms. The results of our studies suggest that the
low-rank algorithms have a substantial performance advantage over
conventional MVDR space-time beamforming algorithms. Furthermore,
the use of sparsity-aware and knowledge-aided strategies is also
able to improve the performance of space-beamforming algorithms, and
can be combined with low-rank schemes. These beamforming algorithms
can be also applied to MIMO radar systems, mono-static and bi-static
radar systems and other sensing applications such as sonar systems.

\bibliographystyle{ws-book-har}    


\printindex

\end{document}